\documentclass[12pt,a4paper]{iopart}
\usepackage{iopams}             
\usepackage{graphicx}           
\usepackage{hyperref}           
\begin{document}

\title[p-Pb collisions at 5.02 TeV in PHSD]
{p-Pb collisions at 5.02 TeV in the Parton-Hadron-String-Dynamics transport approach}

\author{V P Konchakovski$^1$, W Cassing$^1$ and V D Toneev$^2$}
\address{$^1$ Institute for Theoretical Physics, University of Giessen,
  35392 Giessen, Germany}
\address{$^2$ Joint Institute for Nuclear Research,
  141980 Dubna, Russia}

\begin{abstract}
The Parton-Hadron-String-Dynamics (PHSD) transport model is employed
for p-Pb collisions at $\sqrt{s_{NN}}=$ 5.02 TeV and compared to
recent experimental data from the LHC as well as to alternative
models. We focus on the question of initial state dynamics, i.e. if
the initial state might be approximated by a superposition of
independent nucleon-nucleon collisions or should be considered as a
coherent gluon field as predicted within the color glass condensate
(CGC) framework. We find that the PHSD approach provides correlations
between the charged particle multiplicity at midrapidity and the
number of participant nucleons close to results from the CGC and
differs substantially from results calculated with independent Glauber
initial conditions. However, a sizeable difference is found between
the PHSD approach and CGC models with respect to the rapidity
dependence of the average transverse momentum. Accordingly, related
measurements at LHC should allow to prove or disprove the presence of
coherent colour fields in the initial phase of the collisions.
\end{abstract}

\submitto{\JPG}

\maketitle

\section{Introduction}

Ultra-relativistic nucleus-nucleus collisions provide an opportunity
for exploring strongly interacting QCD matter under extreme conditions
which is the ultimate goal of heavy-ion experiments at the
relativistic heavy-ion collider (RHIC) and the large hadron collider
(LHC). The experiments at the RHIC and the LHC have demonstrated that
a stage of partonic matter is produced in these reactions which is in
an approximate equilibrium for a few fm/c~\cite{3years05,QM11}. Due to
the non-perturbative and non-equilibrium nature of relativistic
nuclear reaction systems, their theoretical description is based
essentially on a variety of effective models ranging from hydrodynamic
models with different initial conditions~\cite{IdealHydro1,
  IdealHydro2,IdealHydro3,IdealHydro4,IdealHydro5,
  IdealHydro6,ViscousHydro1,ViscousHydro2,ViscousHydro3,ViscousHydro4}
to various kinetic approaches~\cite{T1,T2,T3,Greco,AMPT,BAMPS,CB09} or
different types of hybrid models
\cite{hyb1,hyb2,hyb3,hyb4,hyb5,hyb6,hyb7} that employ conceptually
different assumptions on the initial conditions. In the latter
approaches the initial state models are followed by an ideal or
viscous hydro phase which after freeze-out is completed by a hadronic
cascade simulation. However, a commonly accepted and complete picture
is still lacking and precise data from the RHIC and LHC are expected
to clarify the situation. The actual questions addressed in this
study are whether the initial state of the colliding nuclei behaves
like a superposition of its constituents or as a coherent gluon field
as predicted in the color glass condensate (CGC)
framework~\cite{GIJV10}. Furthermore, we address the problem of how to
disentangle the different initial state scenarios in final state
observables.

The different phenomenological models that successfully describe
heavy-ion data include coherence effects in the initial state which
can be identified at the level of the wave function and also at the
level of primary particle production. A complete, QCD-based,
dynamical description of the coherence effects is provided within the
color glass condensate concept (cf.\ the reviews~\cite{GIJV10}). Here,
gluon shadowing is taken into account through nonlinear
renormalization group equations, i.e.\ the BK-JIMWLK evolution of
classical Yang-Mills equations that describe gluon fusion at soft
momentum scales. They imply the emergence of a dynamical transverse
momentum scale, the saturation scale $Q_s$, such that gluon modes with
transverse momentum $k_t\leq Q_s(x)$ are in the saturation
regime~\cite{GIJV10}. Such a saturation/suppression of gluon densities
is equivalent to the presence of strong coherent color
fields. Unfortunately, there is no direct experimental observable that
proves the existence of such coherent color fields (or CGC). To
validate the CGC approach, one should compare different observables
calculated within CGC models with alternative approaches that do not
involve the concept of coherent color fields.

Conventional descriptions of ultrarelativistic heavy-ion reactions are
ideal or viscous hydrodynamic models \cite{IdealHydro1,IdealHydro2,
  IdealHydro3,IdealHydro4,IdealHydro5,IdealHydro6,ViscousHydro1,
  ViscousHydro2,ViscousHydro3,ViscousHydro4} or hybrid
approaches~\cite{hyb1,hyb2,hyb3,hyb4,hyb5,hyb6,hyb7} which can be
examined also on an event-by-event basis~\cite{Schenke,Werner}. In
these hydro calculations the initial conditions -- at some finite
starting time of the order of 0.5 fm/c -- have to be evaluated either
in terms of the (standard) Glauber model or other initial state
scenarios like in the IP-glasma model~\cite{Schenke} or the CGC
approach, respectively. Differences between the different initial
state assumptions and dynamical evolutions thus have to be
expected. The applicability of ideal or viscous hydrodynamic models to
proton-nucleus reactions for low multiplicity events, however, is very
much debated. This also holds for hybrid models as long as they employ
a hydro phase.

The flow harmonics $v_n$ have been found to be sensitive to the early
stage of nuclear interaction and in particular their
fluctuations. Indeed, the detailed heavy-ion analysis in
Ref.~\cite{Al11} shows that Monte Carlo CGC approaches (MC-CGC)
systematically give a larger initial eccentricity than Glauber
models. However, it is unclear to what extent such properties of the
CGC formalism are robust with respect to extended correlations. Also,
studies of higher harmonics -- as presented in~\cite{QM11} by the
PHENIX or ALICE collaborations -- do not clearly favor the CGC or
Glauber assumptions for the initial state of the collision. The first
LHC data on the bulk particle production in Pb-Pb collisions are in
good agreement with improved CGC expectations but they are also
compatible with Monte Carlo event
generators~\cite{3years05,ATLAS-fl13}. Both models have in common
that they include some 'coherence effects'.

The complexity of heavy-ion collisions is reduced essentially in the
case of proton-nucleus collisions owing to the expected dominance of
the initial state effects. Recently, the first preliminary ALICE
measurement of the charged particle pseudorapidity density has been
reported~\cite{ALICE:2012xs} for $|\eta|<$~2 in p-Pb collisions at a
nucleon-nucleon center-of-mass energy $\sqrt{s_{NN}}=$ 5.02 TeV. The
measurement is compared to two sets of particle production models that
describe similar measurements for other collision systems: the
saturation models employing coherence
effects~\cite{DK12,TV12,Albacete:2012xq} and the two-component models
combining perturbative QCD processes with soft
interactions~\cite{BBG12,XDW12}. A comparison of the model
calculations with the data shows that the results are model-dependent
and predict the measured multiplicity values only within 20$\%$.
Accordingly, the restrictions imposed by the measured minimal bias
pseudorapidity spectra $dN_c/d \eta$ are not sufficient to disentangle
different models for the very early interaction stage of
ultrarelativistic collisions. A large set of various characteristic
predicted in the compilation~\cite{AAB13} for p-Pb collisions at 5.02
TeV is still waiting for a proper analysis/comparison.

A test of color coherence in proton-nucleus collisions at the LHC
energy has been proposed in Ref.~\cite{BS13}. The idea of this
proposal is based on the fact that the observed mean multiplicity of
charged particles $\left<N_{ch}\right>$ linearly depends on the number
of participants $N_{part}$ within the wounded nucleon model (WNM) of
independent nucleon-nucleon scatterings, $\left<N_{ch}\right>\sim
N_{part}$, while in the CGC models this dependence is logarithmic,
$\left<N_{ch}\right>\sim \ln N_{part}$. For a small number of
participants, $N_{part}\leq$~10, the mean multiplicities calculated in
both approaches practically coincide (in agreement with experiment)
but for $N_{part}\sim$~25 they differ by almost a factor of
two~\cite{BS13}. Such large numbers of participant are possible at the
LHC energy of 5.02 TeV in p-Pb collisions. Furthermore, as pointed out
in Ref.~\cite{BBS13} there should be a sizeable difference in the mean
transverse momentum of particles versus the pseudorapidity
$\left<p_T\right>(\eta)$ with opposite slopes in $\eta$ on the
projectile side within the CGC framework relative to hydrodynamical
calculations due the saturation scale $Q_s$ in the CGC.

Following these suggestions we here study the charged particle
multiplicities and related quantities in p-Pb interactions at the
collision energy $\sqrt{s_{NN}}=$~5.02 TeV within the
parton-hadron-string-dynamics (PHSD) transport approach~\cite{CB09}
which has been properly upgraded to LHC energies with respect to a
more recent PYTHIA implementation (Sec.~II). Predictions for various
observables and their correlations are given in Sec.~III which are
also compared to available data as well as to results from CGC
saturation models. We conclude our findings in Sec.~IV.

\section{PHSD @ LHC}

The PHSD model is a covariant dynamical approach for strongly
interacting systems formulated on the basis of Kadanoff-Baym
equations~\cite{JCG04} or off-shell transport equations in phase-space
representation, respectively. In the Kadanoff-Baym theory the field
quanta are described in terms of dressed propagators with complex
selfenergies. Whereas the real part of the selfenergies can be related
to mean-field potentials (of Lorentz scalar, vector or tensor type),
the imaginary parts provide information about the lifetime and/or
reaction rates of time-like particles~\cite{Ca09}. Once the proper
(complex) selfenergies of the degrees of freedom are known, the time
evolution of the system is fully governed by off-shell transport
equations (as described in Refs.~\cite{JCG04,Ca09}). This approach
allows for a simple and transparent interpretation of lattice QCD
results for thermodynamic quantities as well as correlators and leads
to effective strongly interacting partonic quasiparticles with broad
spectral functions. For a review on off-shell transport theory we
refer the reader to Ref.~\cite{Ca09}; model results and their
comparison with experimental observables for heavy-ion collisions from
the lower super-proton-synchrotron (SPS) to RHIC energies can be found
in Refs.~\cite{CB09,To12,KB12} including electromagnetic probes such
as $e^+e^-$ or $\mu^+\mu^-$ pairs~\cite{el-m}. We mention that the
PHSD model takes into account some kind of 'coherent effects' with
respect to QCD showers since it includes corrections to the
leading-log picture -- denoted as coherence effects -- that lead to an
ordering of subsequent emissions in terms of decreasing angles.

\begin{figure*}[th]
\centering
\includegraphics[width=0.49\textwidth]{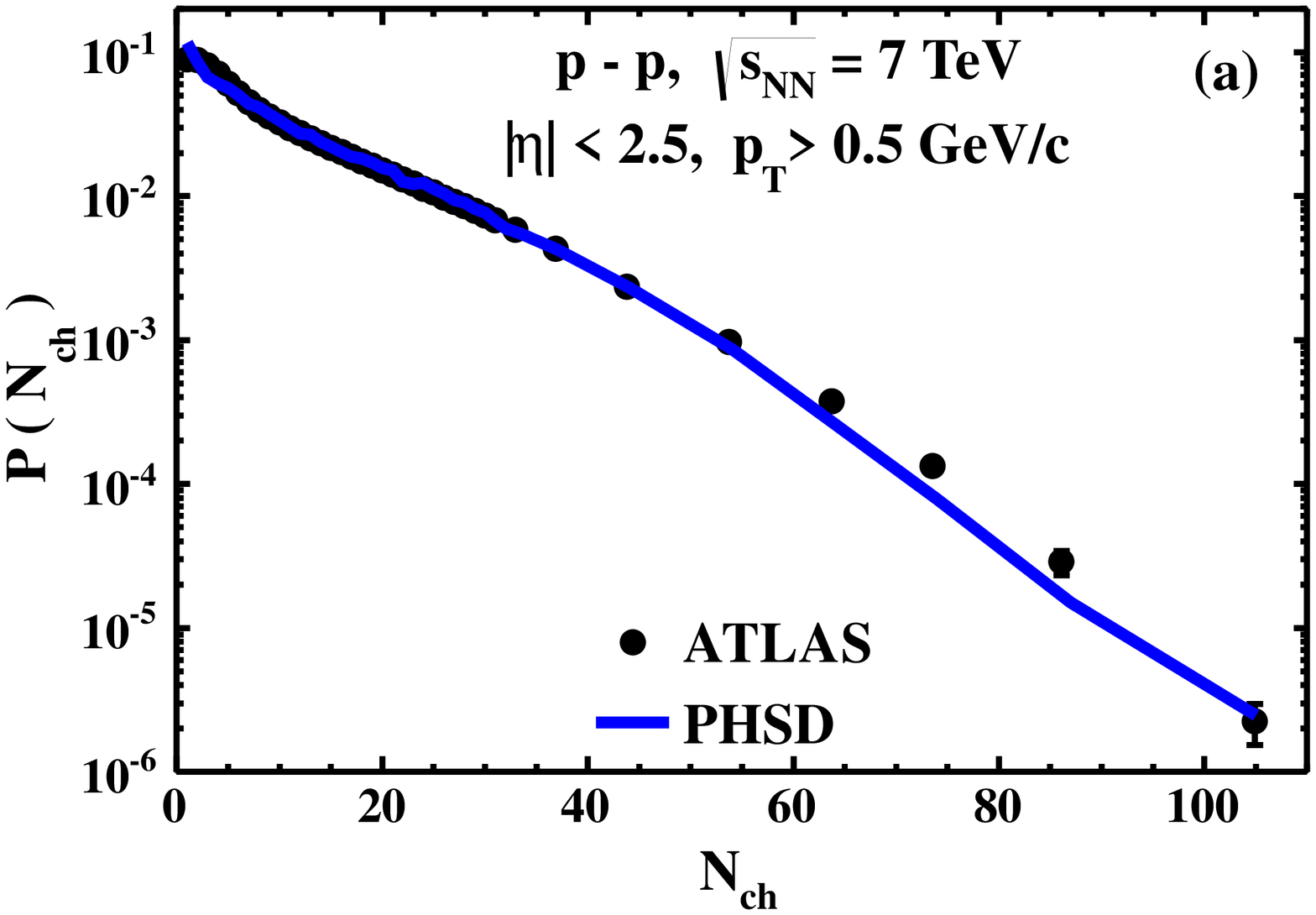}
\includegraphics[width=0.49\textwidth]{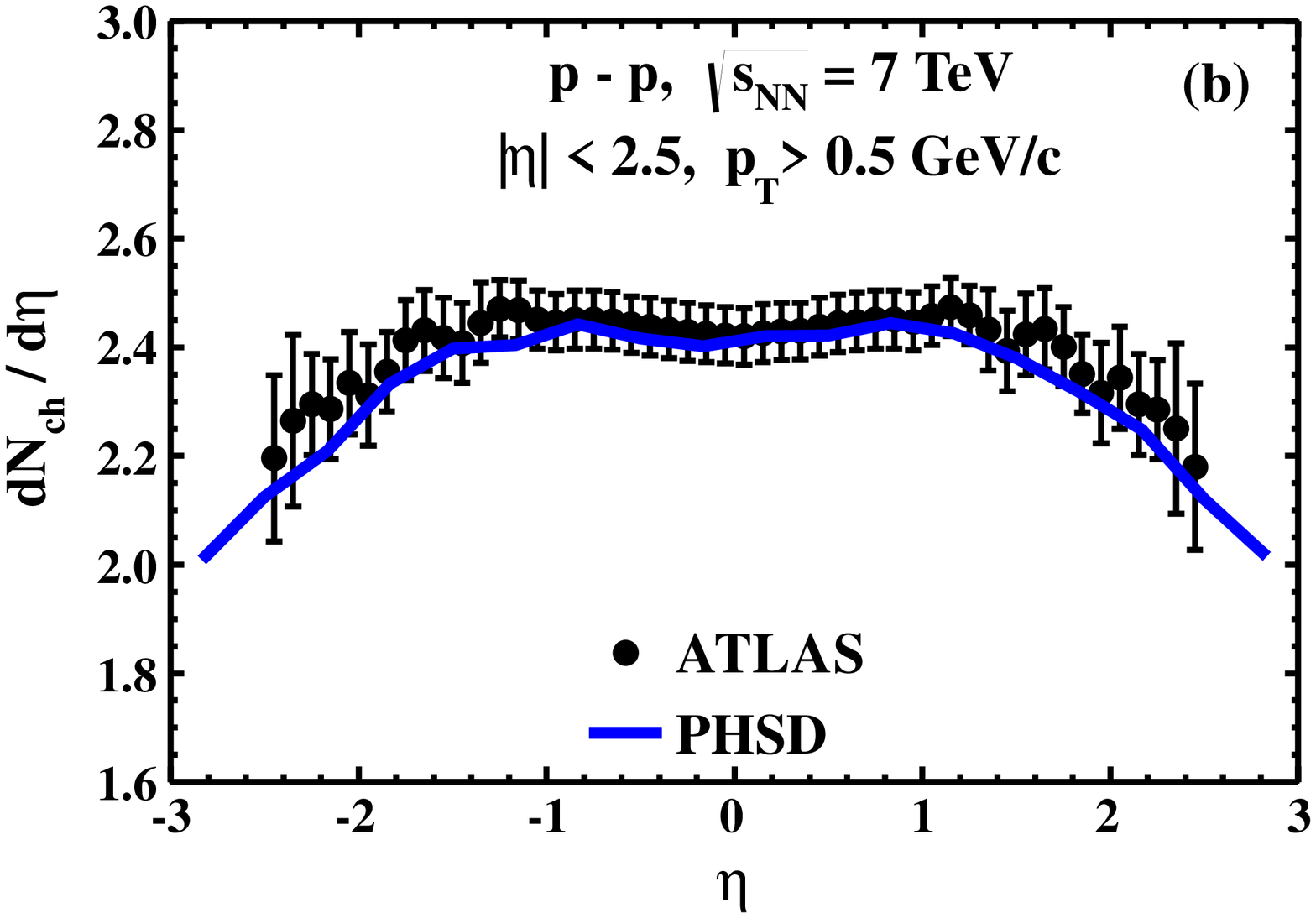}
\includegraphics[width=0.49\textwidth]{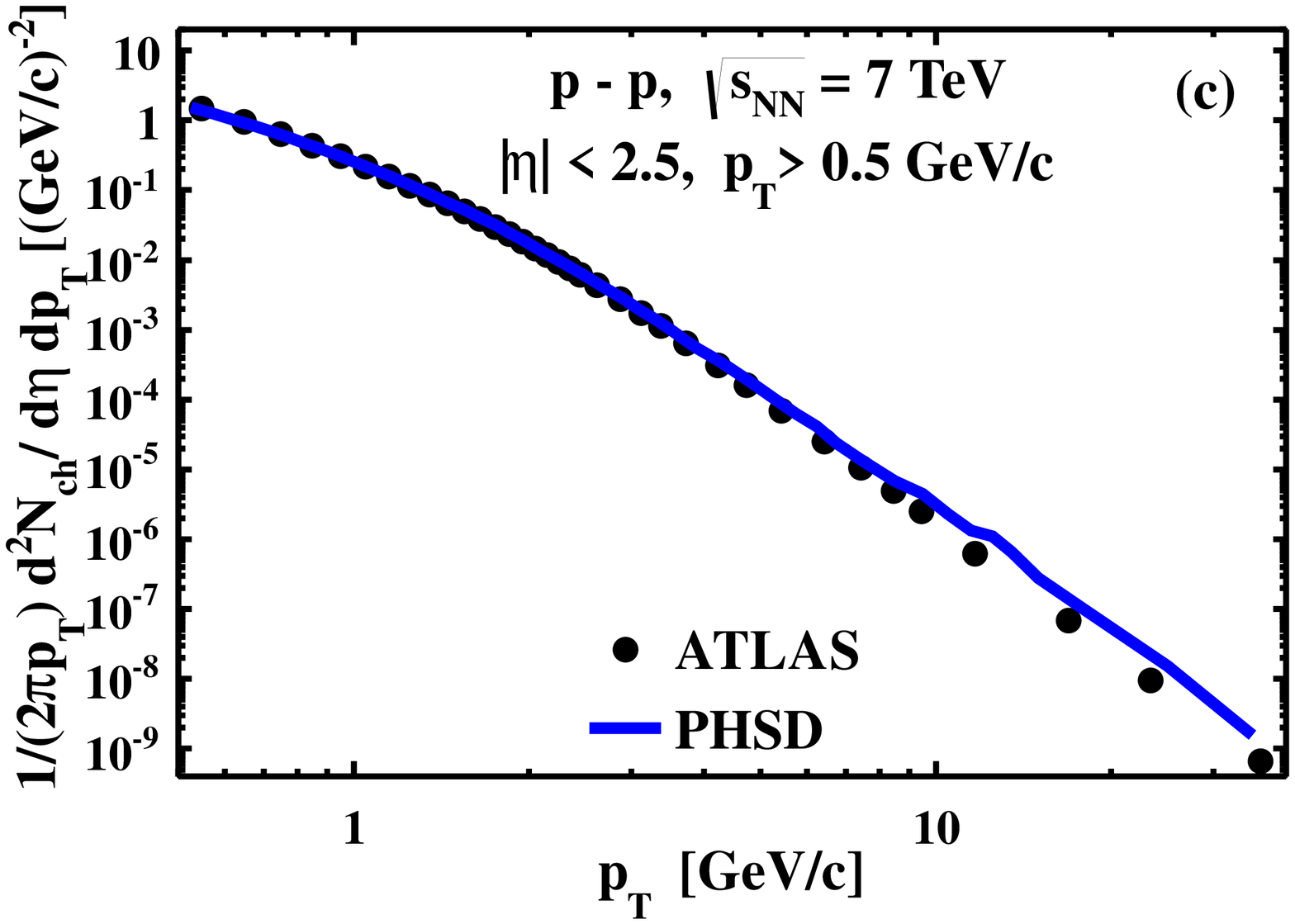}
\includegraphics[width=0.49\textwidth]{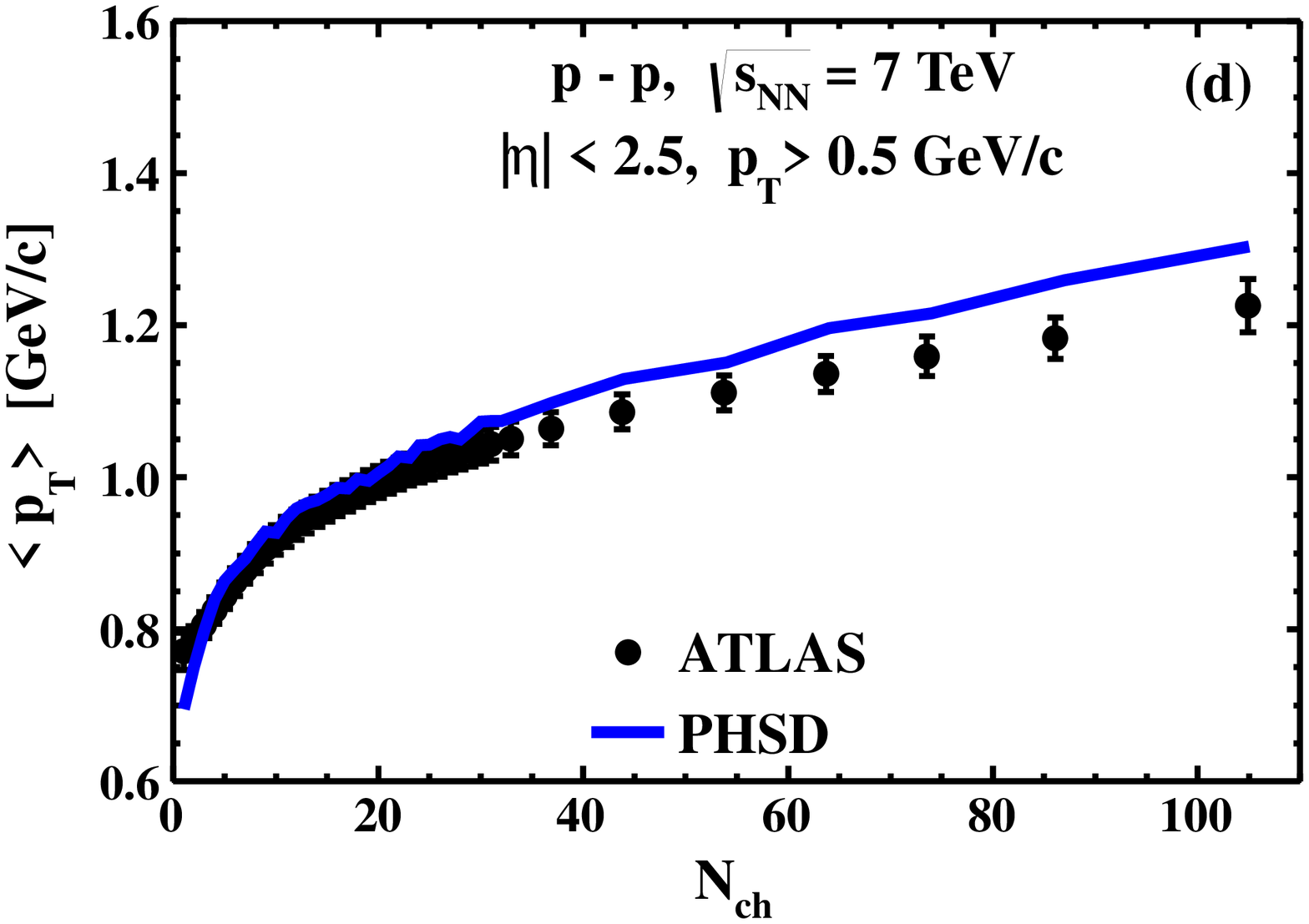}
\caption{Comparison of the PHSD results (including PYTHIA 6.4) with
  LHC experimental data from the ATLAS Collaboration~\cite{Aad:2010ac}
  for p-p collisions at $\sqrt{s_{NN}}=$~7 TeV: (a) $N_{ch}$
  distribution, (b) $dN_{ch}/d\eta$ distribution, (c) $p_T$-spectra
  and (d) average $p_T$ vs.\ $N_{ch}$.}
\label{fig:pp}
\end{figure*}

To extend the PHSD model to higher energies than $\sqrt{s_{NN}}=
200$~GeV at RHIC, we have additionally implemented the PYTHIA 6.4
generator~\cite{Sjostrand:2006za} for initial nucleon collisions at
LHC energies. For the subsequent (lower energy) collisions the
standard PHSD model~\cite{CB09} is applied (including PYTHIA v5.5 with
JETSET v7.3 for the production and fragmentation of
jets~\cite{PYTHIA0}, i.e.\ for $\sqrt{s_{NN}} \le $ 500
GeV~\cite{PYTHIA0}). In this way all results from PHSD up to top RHIC
energies are regained and a proper extension to LHC energies is
achieved. At $\sim \sqrt{s_{NN}} = $ 500 GeV both PYTHIA versions lead
to very similar results. In PYTHIA 6.4 we use the Innsbruck pp tune
(390) which allows to describe reasonably the p-p collisions at
$\sqrt{s_{NN}} =$ 7 TeV in the framework of the PHSD transport
approach (cf.\ Fig.~\ref{fig:pp}). Here the overall agreement with LHC
experimental data for the distribution in the charged particle
multiplicity $N_{ch}$ (a), the charged particle pseudorapidity
distribution (b), the transverse momentum $p_T$ spectra (c) and the
correlation of the average $p_T$ with the number of charged particles
(d) is satisfactory. One should note that the experimental results of
different LHC collaborations slightly differ (within errorbars). In
particular, the pseudorapidity and transverse momentum distributions
of the CMS Collaboration are slightly below those of the ATLAS
Collaboration. In addition, different PYTHIA tunes, being generally in
satisfactory agreement with pp data and tuned to specific observables,
describe the tails of $N_{ch}$ and $p_T$ distributions with different
quality.

Although PYTHIA 6.4 includes some elements of coherence in the
creation of particles (by string fusion, string fragmentation,
quasiparticle spectral densities etc.) it deviates substantially from
the early interaction stage in the CGC approach~\cite{GIJV10}. We
mention that initial state fluctuations in hydro calculations are
usually imposed by independent Glauber model simulations or MC-KLN
initial conditions~\cite{Luzum14}, respectively. Initial conditions
very similar to the Glauber model are included by default in the PHSD
transport approach, however, with an essential difference: the
energy-momentum conservation is fulfilled exactly in every collision
such that the entire dynamics conserves four-momentum as well as all
discrete conservation laws. We recall that the PHSD approach has been
tested successfully for collective flows $v_1, v_2, v_3$ and $v_4$ in
nucleus-nucleus collisions from lower super-proton-synchrotron (SPS)
up to RHIC energies~\cite{To12} where especially the uneven flow
coefficients are sensitive to the initial state fluctuations.
Accordingly, the initial fluctuations in energy density - in the
transverse plane - from PHSD are in accord with experimental
observation.

It has been argued, furthermore, that in high energy p-A collisions
the Glauber model should be corrected/extended to account for the fact
that between successive interactions the incoming proton is
off-shell~\cite{Gr69} and may fluctuate in size. In addition,
event-to-event fluctuations in the configuration of the incoming
proton can change its effective scattering cross section as noted in
Refs.~\cite{HBB91,GS06,AS13}. The concept of hadronic cross-section
fluctuations incorporates the physics of color transparency and color
opacity into the dynamics of relativistic nuclear collisions. In
order to evaluate the impact of these fluctuations of the projectile
proton, a modified version of the Glauber Monte Carlo, referred to as
'Glauber-Gribov' MC, is implemented additionally/optionally in the
PHSD. Following Refs.~\cite{HBB91,GS06,AS13}, the probability
distribution in the total cross sections $\sigma_{tot}$ is taken to be
\begin{eqnarray}
P_h(\sigma_{tot})=a_h \frac{\sigma_{tot}}{\sigma_{tot}+\sigma_0}
\exp{\left(-\frac{(\sigma_{tot}/\sigma_0-1)^2}{\Omega^2}\right)}~.
\label{GGsigma}
\end{eqnarray}
Here, $a_h$ is a normalization constant, $\Omega$ controls the width
of the $P_h(\sigma_{tot})$ distribution, and $\sigma_0$ determines
$\left<\sigma_{tot}(\sqrt{s})\right>$ which is adopted from PYTHIA
6.4. Estimates of $\Omega$ have been provided in Ref.~\cite{AS13} for
center-of-mass energies of 1.8, 9, and 14 TeV. We use two
interpolations of these values that for $\sqrt{s_{NN}}=$~5.02 TeV
results in $\Omega=$~0.55 or, in accordance with the recent analysis
in Ref.~\cite{AS13}, to $\Omega=$~1.01 which enhances the influence of
fluctuations. The elastic fraction of the total cross-section
in~(\ref{GGsigma}) is taken to be constant~\cite{AS13}, $\sigma_{NN} =
\lambda \sigma_{tot}$ (following~\cite{AS13} $\lambda$ is weakly
changing with energy and the actual value employed is $\lambda=0.26$);
the probability distribution for $\sigma_{NN}$ then is given by
$P_H(\sigma_{NN}) =(1/\lambda) P(\sigma_{NN}/\lambda)$. The actual
values for the elastic and inelastic cross sections in PHSD are
determined by Monte Carlo according to the distribution
(\ref{GGsigma}).

\section{Properties of p-Pb collisions}

\subsection{Energy density in a single p-Pb event}

\begin{figure}[th]
\centering
\includegraphics[width=0.49\textwidth]{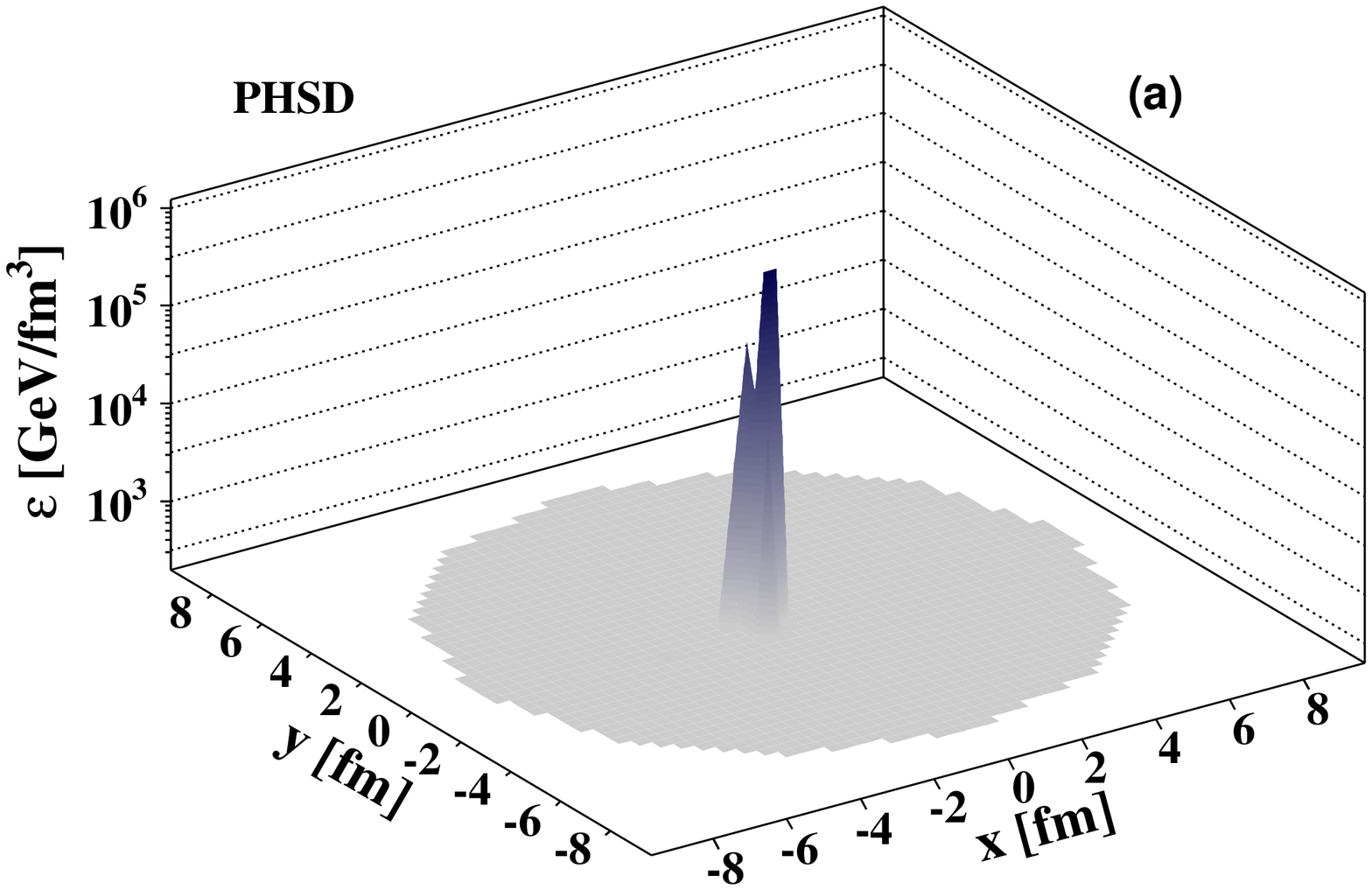}
\includegraphics[width=0.49\textwidth]{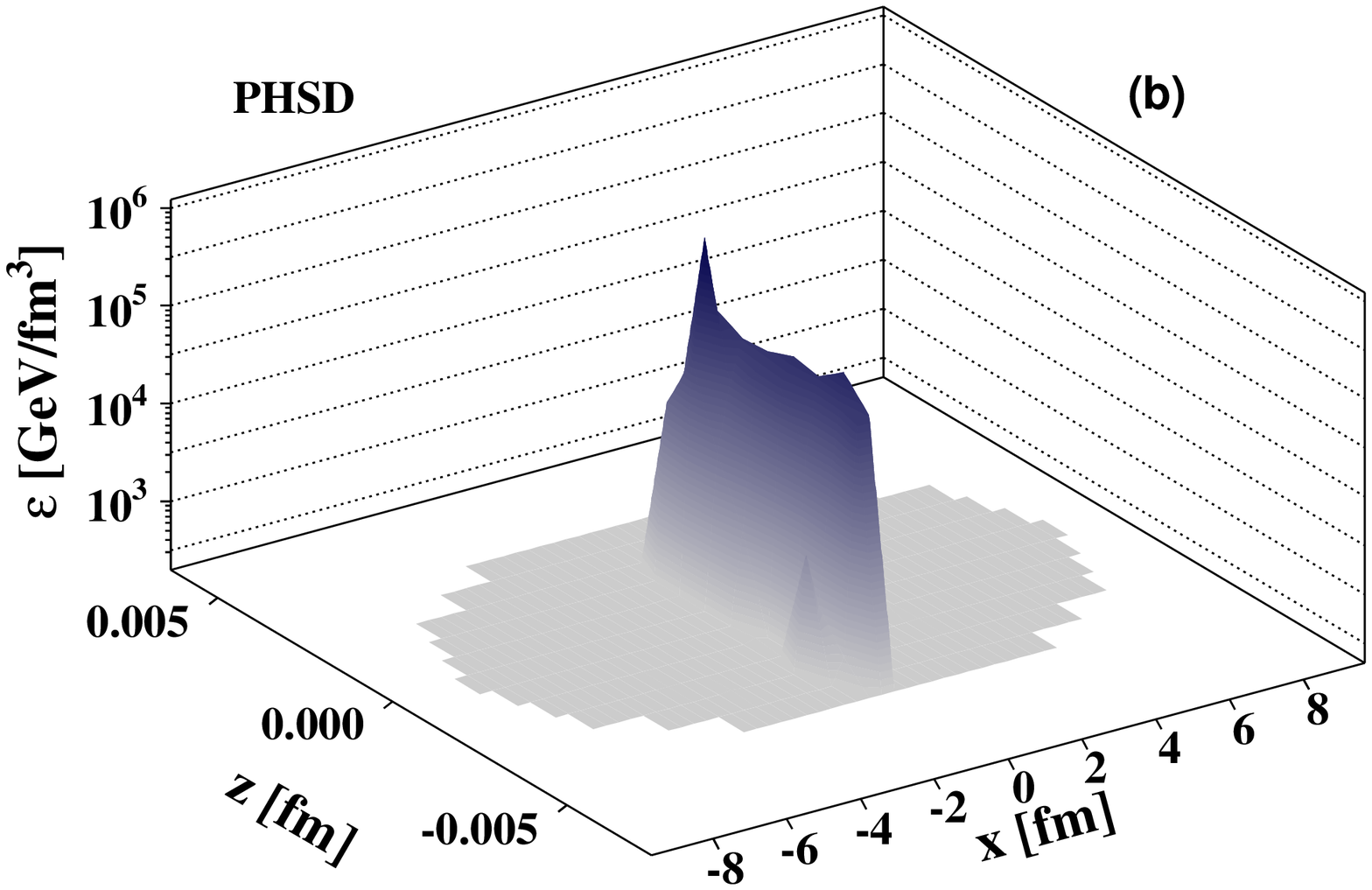}
\caption{$x-y$ (a) and $x-z$ (b) projections of the energy density in
  a single (highly inelastic) p-Pb event ($\sqrt{s_{NN}}$ = 5.02 TeV)
  at the time when the proton-remnant has passed through the
  Pb-nucleus ($t=$~0.002 fm/c). The region occupied by the Pb nucleus
  is also shown by the shaded area. Note the Lorentz contracted scale
  in $z$-direction.}
\label{fig:edens}
\end{figure}

The energy density $\epsilon$ in local cells from PHSD is presented in
Fig.~\ref{fig:edens} in the transverse ($x-y$) plane (a) as well as
the reaction ($x-z$) plane (b) for a single p-Pb event at
$\sqrt{s_{NN}}=$ 5.02 TeV. Here we have selected an event that leads
to about 300 charged hadrons in the final state. The time of this
event ($t=$~0.002 fm/c) corresponds to the moment when the proton has
passed the Lorentz contracted nucleus and is very small compared to
the initial times considered for hydrodynamical models ($\sim$ 0.5 - 1
fm/c). Note that the energy density at this moment is huge due to the
fact that the spacial volume is very tiny ($\sim 5 \cdot 10^{-3}$
fm$^3$) and the full inelasticity of the previous inelastic reactions
is incorporated. Due to the Heisenberg uncertainty relation this
energy density cannot be specified as being due to 'particles' since
the latter may form only much later on a timescale of their inverse
transverse mass (in their rest frame). More specifically, only a jet
at midrapidity with transverse momentum $p_T =$ 100 GeV is expected to
appear at $t \approx 2\cdot 10^{-3}$ fm/c while a soft parton with
transverse momentum $p_T = 0.5$ GeV should be formed after $t \approx$
0.4 fm/c). At this time the energy density $\epsilon$ is lower by more
than a factor of 200 due to the dominant longitudinal expansion.

The maximal energy density in the p-Pb reaction (at $t=$~0.002 fm/c)
is comparable with that in heavy-ion collisions at the LHC energy for
local cells due to fluctuations of the initial conditions in PHSD (in
case of a high spatial resolution). Note that the formed high energy
density "tube"\ is strongly Lorentz contracted along the collision
axis $z$ and is reminiscent of the energy density in a 'string' that
stretches in the longitudinal direction with increasing time.

\subsection{Charged particle multiplicities and their distributions}

\begin{figure}[!b]
\centering
\includegraphics[width=0.49\textwidth]{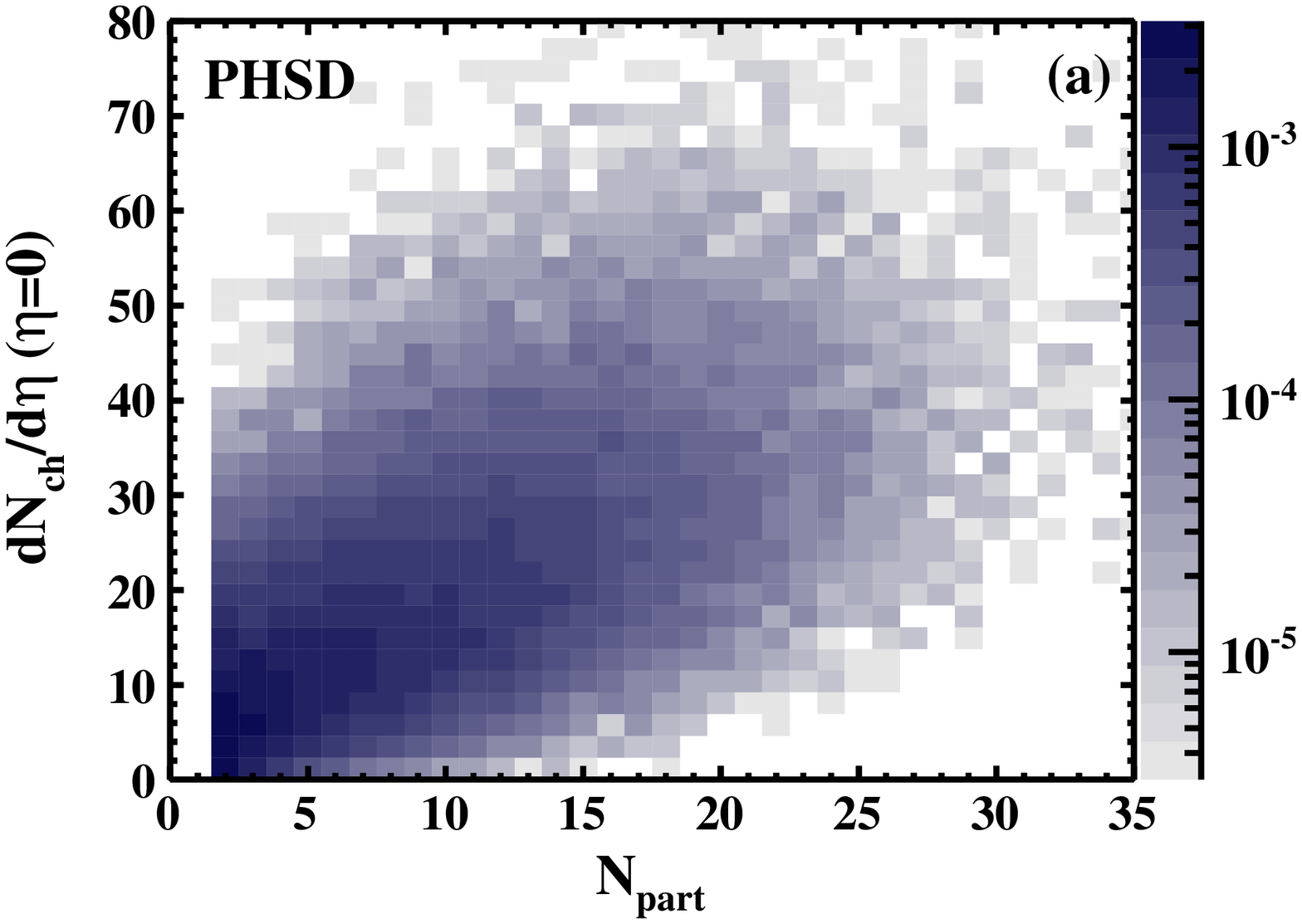}
\includegraphics[width=0.49\textwidth]{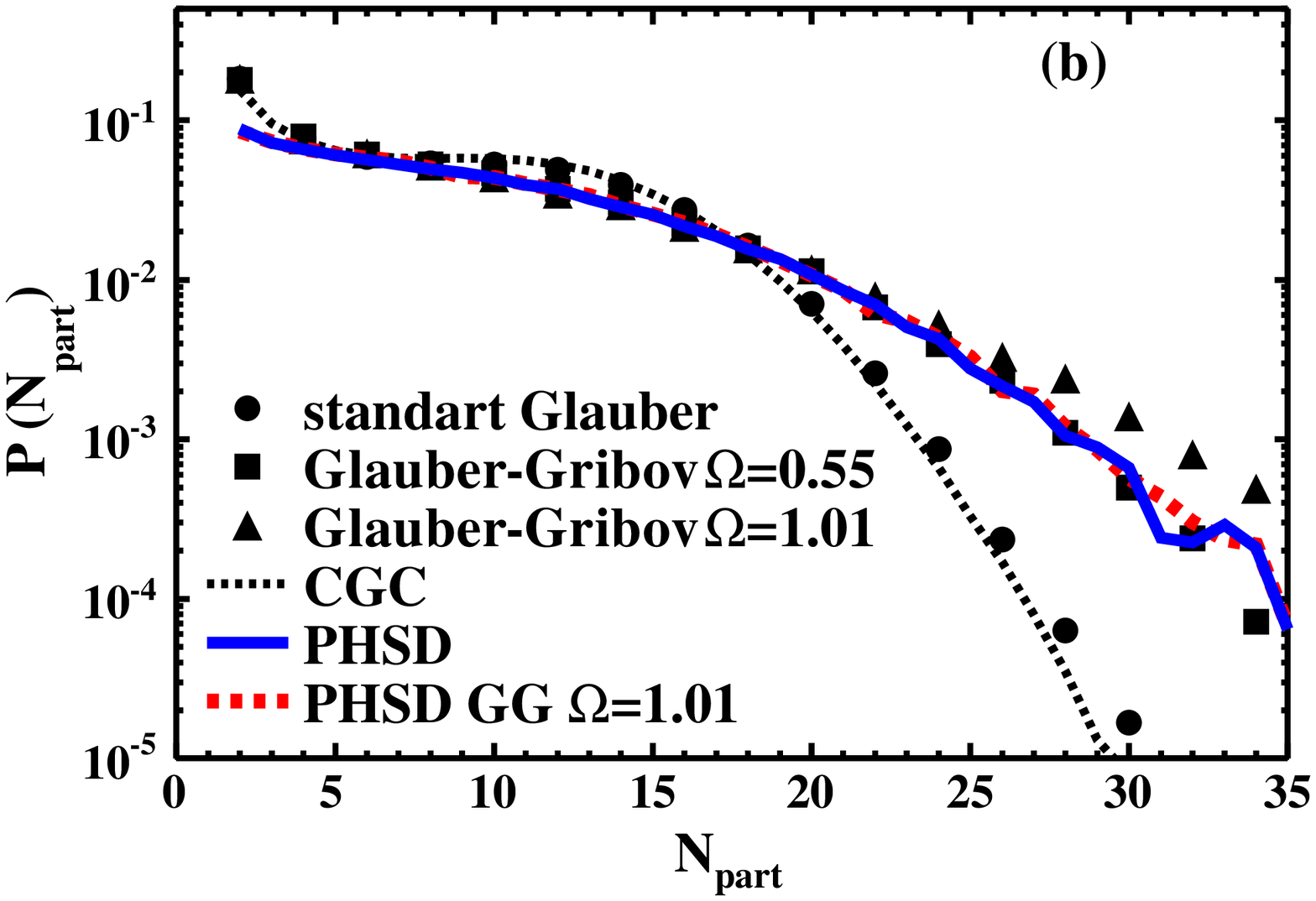}
\includegraphics[width=0.49\textwidth]{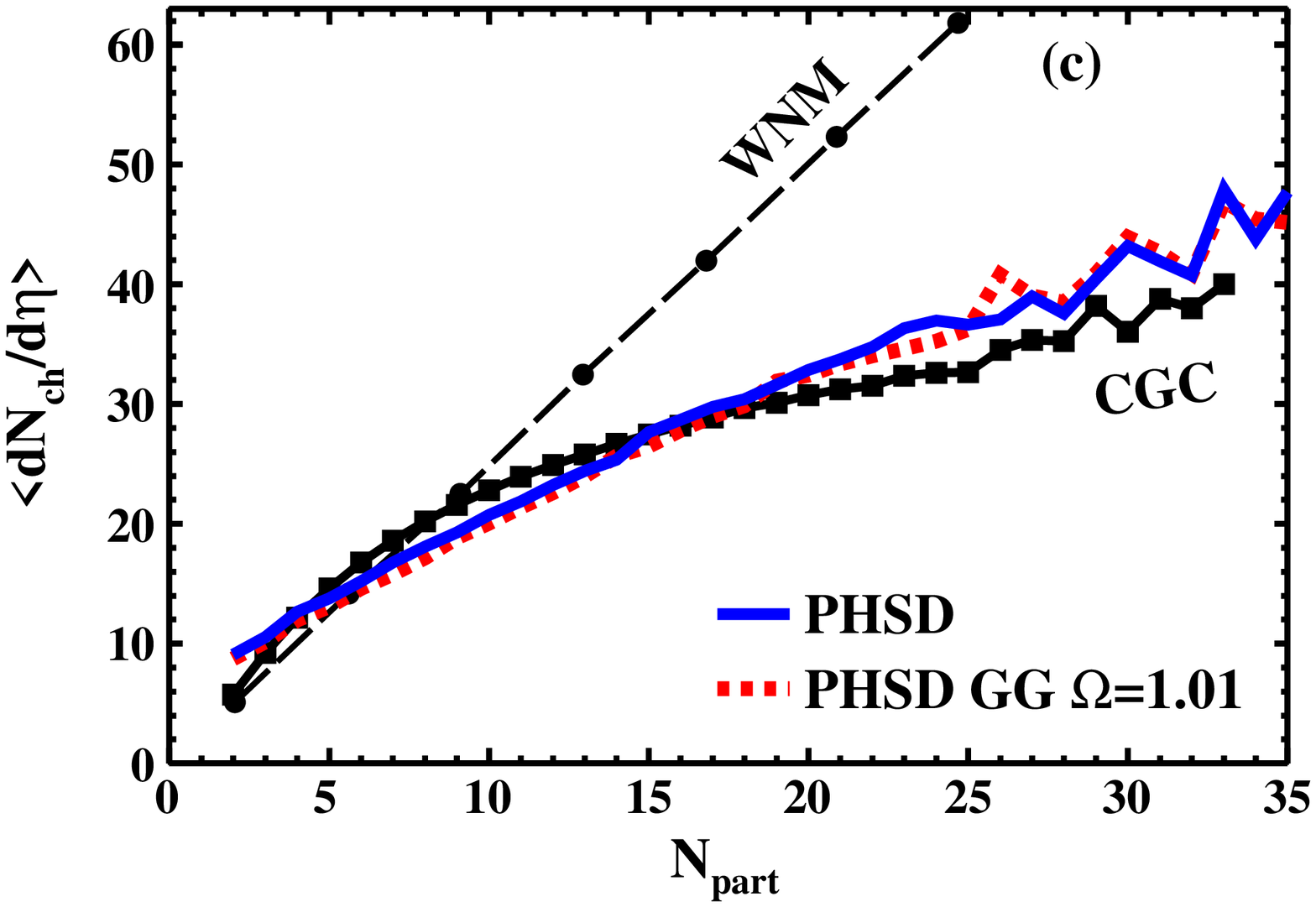}
\caption{Probability distribution of the participant number and number
  of charged particles for p-Pb at $\sqrt{s_{NN}}$ = 5.02 TeV at
  midrapidity (a) and its different projections in (b) and (c). The
  wounded nucleon model (WNM)(full dots -standart Glauber) and color
  glass condensate (CGC) calculations (dotted line in (b)) are taken
  from~\cite{BS13} while simulations in the Glauber-Gribov
  approximation (full squares and triangles) stem
  from~\cite{TheATLAScollaboration:2013cja}. The PHSD results are
  displayed in terms of the solid (blue) lines while the PHSD results
  including fluctions in the cross section are shown in terms of the
  dotted (red) lines.}
\label{fig:gribov}
\end{figure}

\begin{figure*}[th]
\centering
\includegraphics[width=0.49\textwidth]{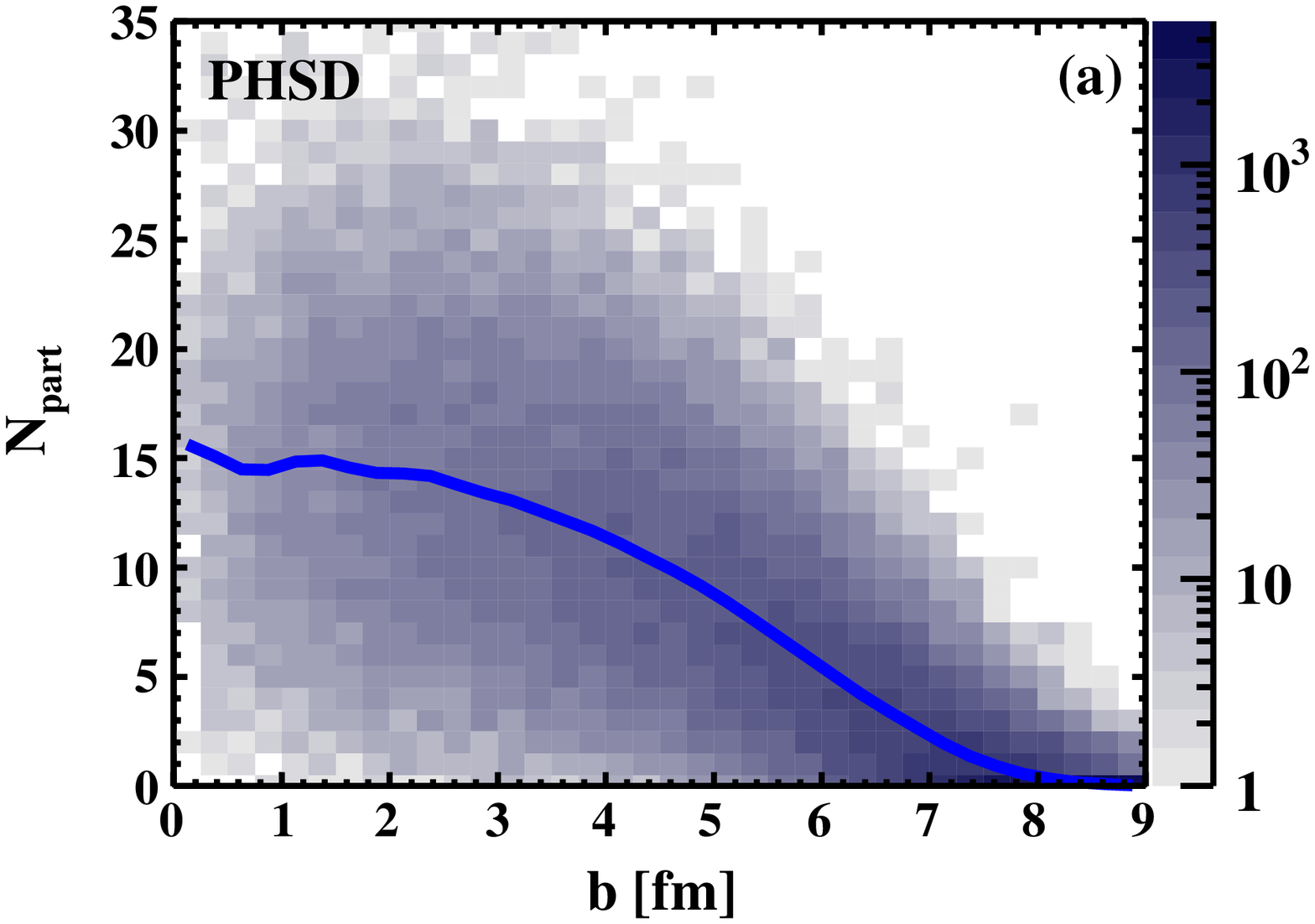}
\includegraphics[width=0.49\textwidth]{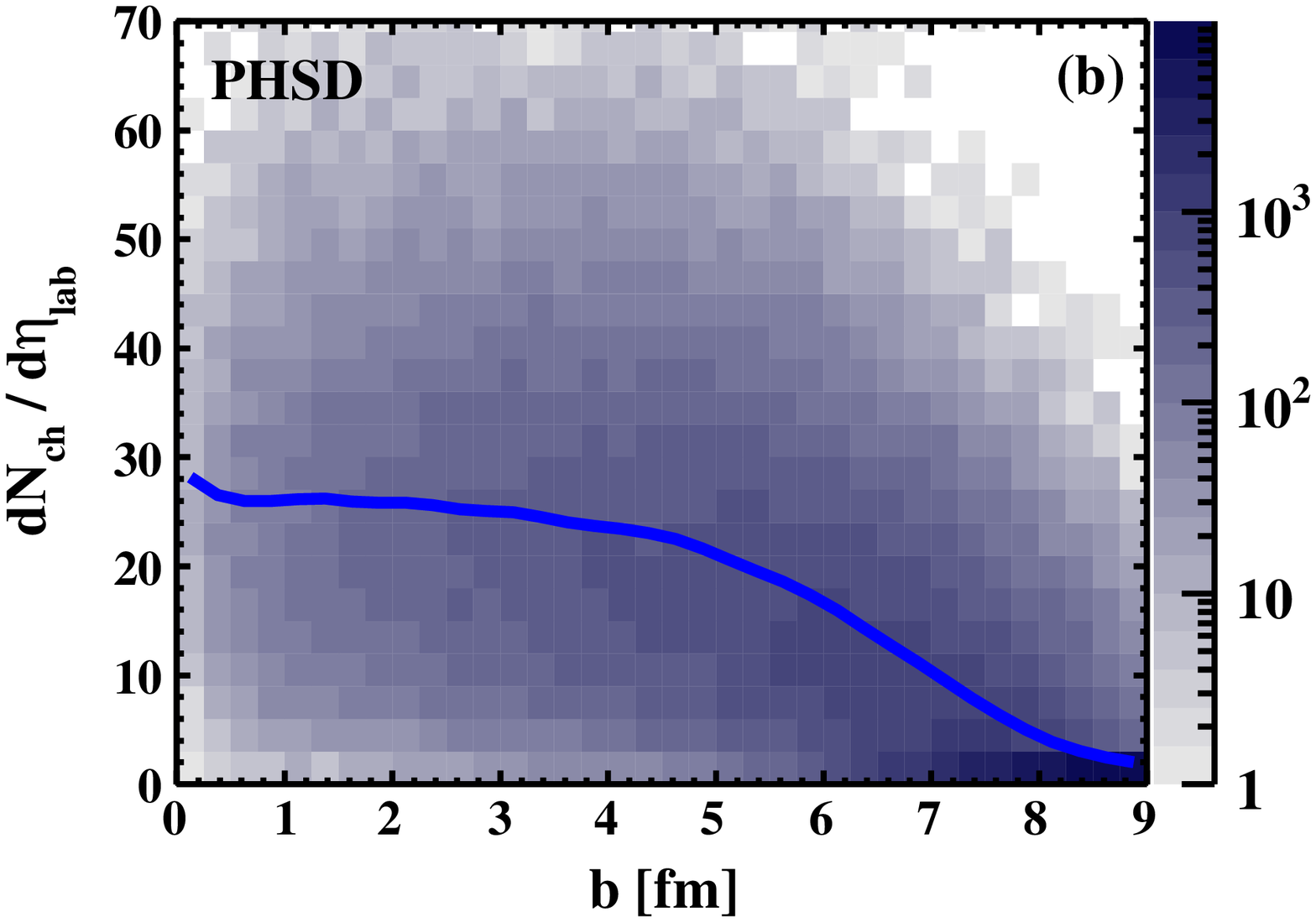}
\caption{Event distributions of 2D-correlations for the participant
  number $N_{part}$ (a) and charged particle multiplicity
  $dN_{ch}/d\eta$ (b) with the impact parameter $b$. The mean values
  of these distributions are shown by the solid (blue) lines.}
\label{fig:bimp}
\end{figure*}

With the elementary p-p collisions in the PHSD being adjusted at LHC
energies via PYTHIA 6.4 (using the Innsbruck pp tune (390)) we now
proceed with observables and correlations from p-Pb collisions. In
Fig.~\ref{fig:gribov}(a) we present the probability distribution in
the participant number $N_{part}$ and the number of charged hadrons at
midrapidity $N_{ch}(\eta=0)$ as well as the different projections for
p-Pb (5.02 TeV) in (b) and (c). In this figure it was assumed that the
charged particles are distributed according to negative binomial
distributions in the Glauber calculation. As is seen from
Fig.~\ref{fig:gribov}, the number of charged particles at midrapidity
correlates with the number of participants, $N_{ch}(\eta=0)\sim
N_{part}$, however, with a large dispersion in both quantities. If
this 2D distribution is integrated over the number of charged
particles, the $P(N_{part})$ distributions for various models are
compared in Fig.~\ref{fig:gribov}(b) and~\ref{fig:gribov}(c). For
$N_{part}\gtrsim$~15, the Gribov-Glauber (GG) distributions
(calculated in the WNM) (full squares and triangles in (b))
increasingly overshoot the standard Glauber (G) result (full dots in
(b))\footnote{In contrast to the PHSD-GG case, for the Glauber
  calculations we use the constant parameter $\Omega=$~0.55 or
  $\Omega=$~1.01 to enhance the influence of cross section
  fluctuations.} and this difference reaches an order of magnitude in
the case of $N_{part}\gtrsim$~30 while all evaluated distributions
practically coincide for low numbers of participants,
$N_{part}\lesssim$~15. In contrast to the Glauber or Gribov-Glauber
Monte Carlo simulations we find no dramatic enhancement in the
distribution when taking into account the cross section fluctuations
in the PHSD (PHSD-GG, red dotted line compared to the blue solid
line); the $P(N_{part})$ distribution is close to the Glauber-Gribov
results in PHSD (Fig.~\ref{fig:gribov}(b)). The noted difference is
seen in the correlation $N_{ch}/d \eta (\eta=0)$ vs.\ $N_{part}$.
Both the standard Glauber and CGC results are presented and support
the results of Ref.~\cite{BS13}. However, the two versions of the PHSD
model, with (red dotted line) and without cross section fluctuations
(blue solid line), predict that the multiplicity dependence turns out
to be close to the CGC result and is only weakly sensitive to the
parameter $\Omega$ for the size of the fluctuations in the cross
section. Thus, multiplicity distributions do not allow us to
disentangle the different initial states under discussion. The reason
of such a multiplicity suppression is the energy-momentum conservation
in PHSD which on average results in a decrease of particle
multiplicity in subsequent scatterings as compared to the primary
interaction. This is directly confirmed by a degradation of the energy
density distribution in the longitudinal direction in
Fig.~\ref{fig:edens}(b).

A wide distribution is also observed in the number of participants or
in the number of charged particles at midrapidity for a given impact
parameter $b$, see Fig.~\ref{fig:bimp}. The solid lines in this figure
show the mean values $\left<N_{part}\right>$ and
$\left<dN_{ch}/d\eta\right>$, respectively. In contrast to
nucleus-nucleus collisions, these mean quantities are almost
independent of the impact parameter for central and semi-central
collisions, $b\lesssim$~(4-5) fm. This fact should have been expected
since the size of the projectile-proton is noticeably smaller than
that of the target nucleus. From these results it follows that an
event selection with respect to the number of participants or charged
particles refers to a large range in impact parameter.

\subsection{Rapidity distribution}

\begin{figure*}[th]
\centering
\includegraphics[width=0.6\textwidth]{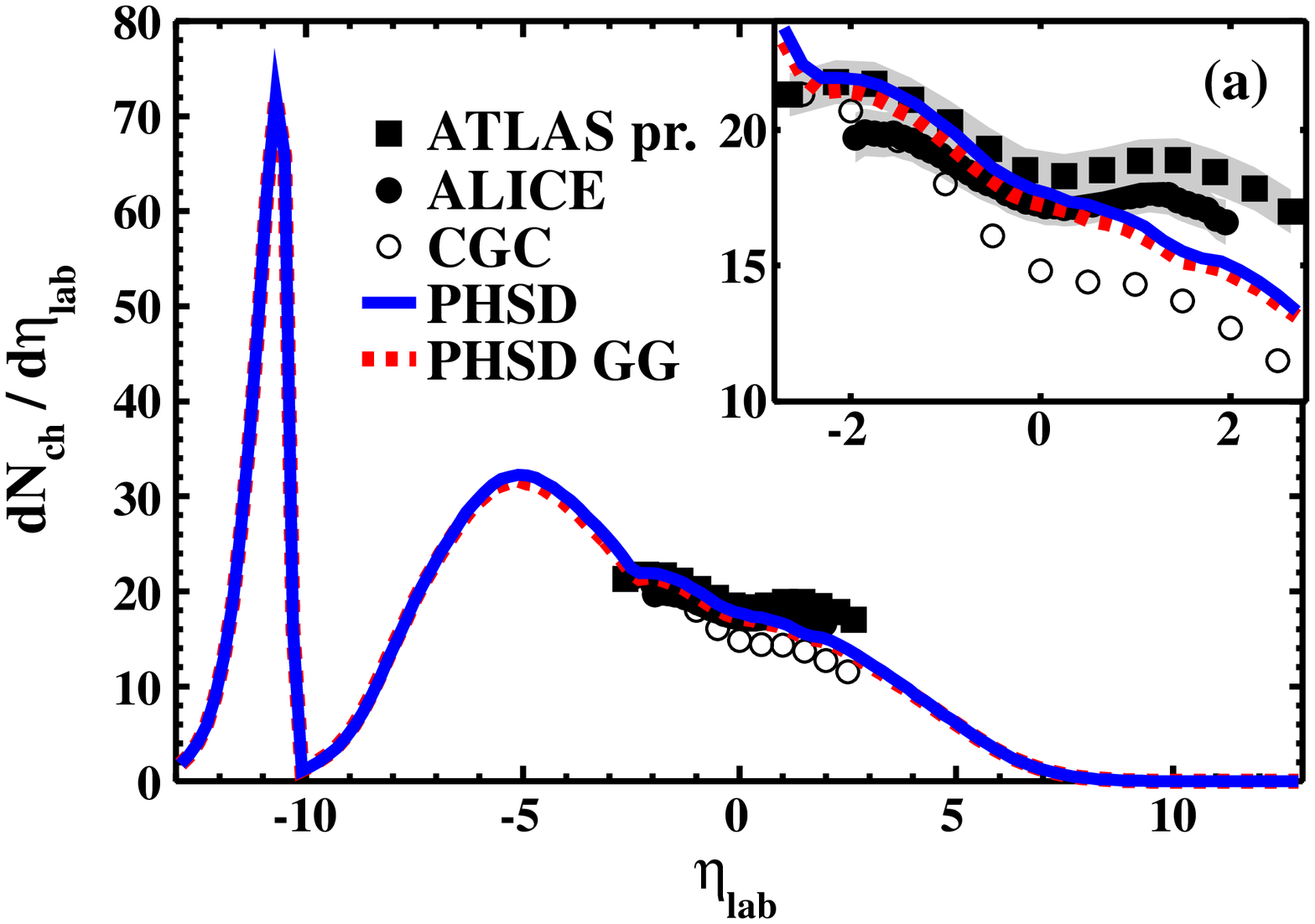}
\includegraphics[width=0.6\textwidth]{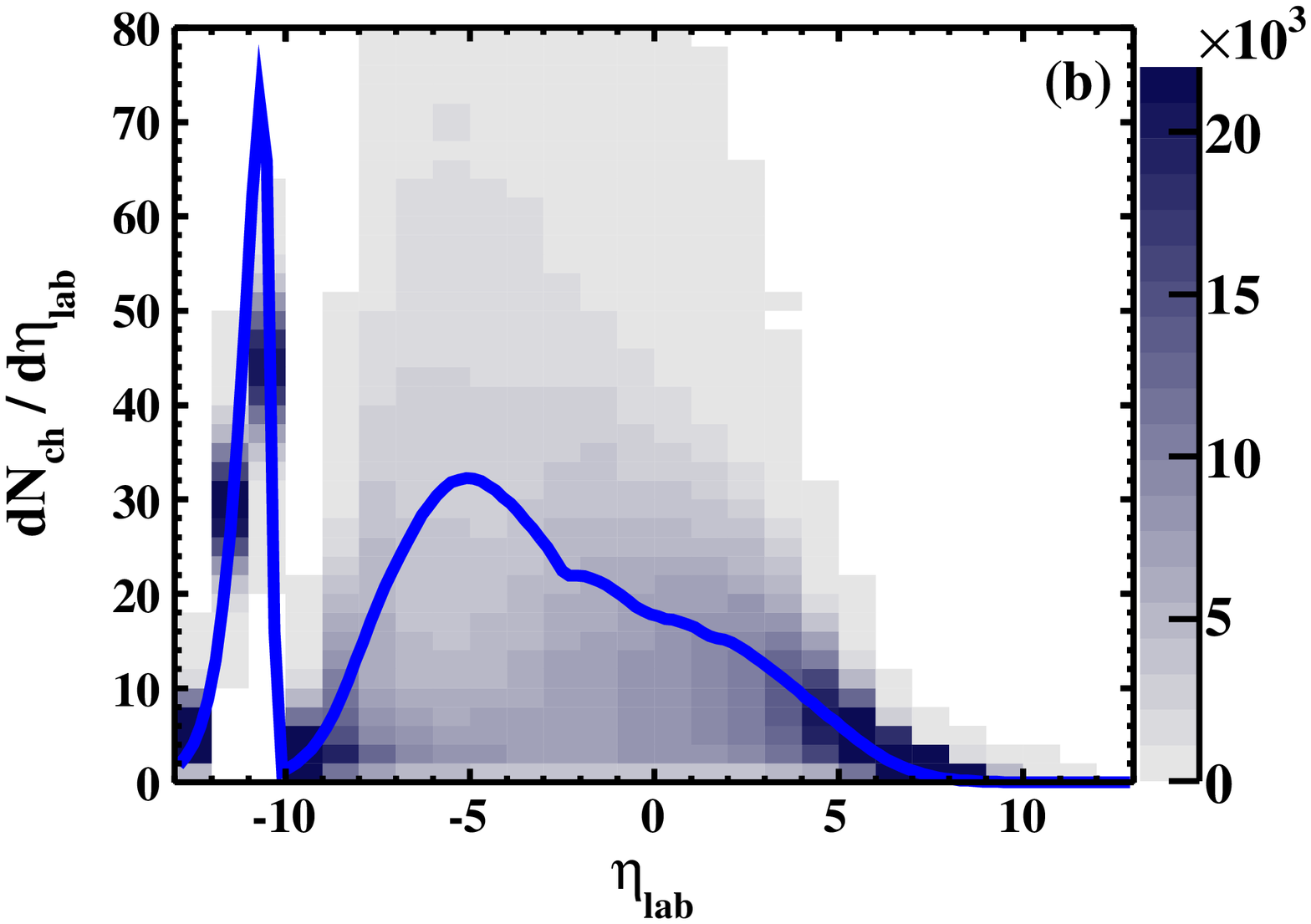}
\caption{(a) Rapidity distribution of charged particles for minimum
  bias data from the ALICE~\cite{ALICE:2012xs} (full dots) and
  ATLAS~\cite{TheATLAScollaboration:2013cja} (full squares)
  collaborations for p-Pb collisions at $\sqrt{s_{NN}}=$~5.02 TeV in
  comparison to the PHSD results (solid blue line) and the PHSD GG
  results including fluctuations in the cross section (dotted red
  line). The CGC results (open circles) have been taken from
  Ref.~\cite{Albacete:2012xq}. The zoomed results are displayed in the
  insertion. (b) Event-by-event fluctuations of the rapidity
  distribution. The blue solid line shows the average charged particle
  pseudorapidity distribution.}
\label{fig:dndeta}
\end{figure*}

The pseudorapidty distributions of charged particles from p-Pb minimum
bias collisions at $\sqrt{s_{NN}}=$~5.02 TeV are compared with the
experimental data~\cite{ALICE:2012xs} in Fig.~\ref{fig:dndeta}. The
data are displayed in the laboratory system which is shifted with
respect to the nucleon-nucleon center-of-mass by $y_{cm}=-0.465$. The
results of two versions of the parton-hadron string dynamics model
(PHSD and PHSD-GG) differ only for backward-emitted particles and both
versions are rather close to the measured data and the CGC result
(open circles). Note that there are no modifications (or free
parameters) in the PHSD except the extensions pointed out in Sec.~II
which implies that p-p, p-A and A-A collisions are consistently
described from low SPS to LHC energies (within $\sim$ 10\%).

\begin{figure*}[th]
\centering
\includegraphics[width=0.6\textwidth]{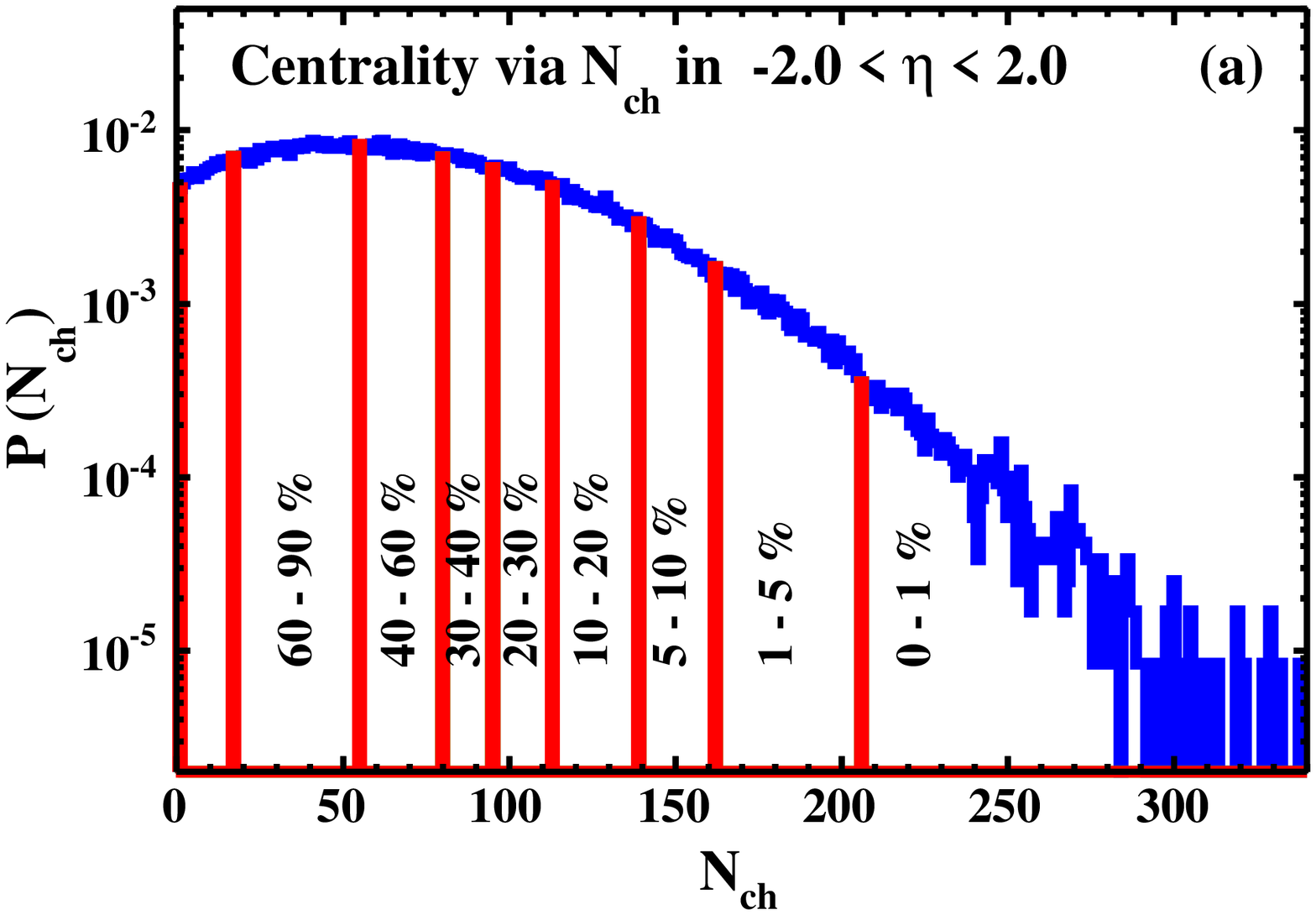}
\includegraphics[width=0.6\textwidth]{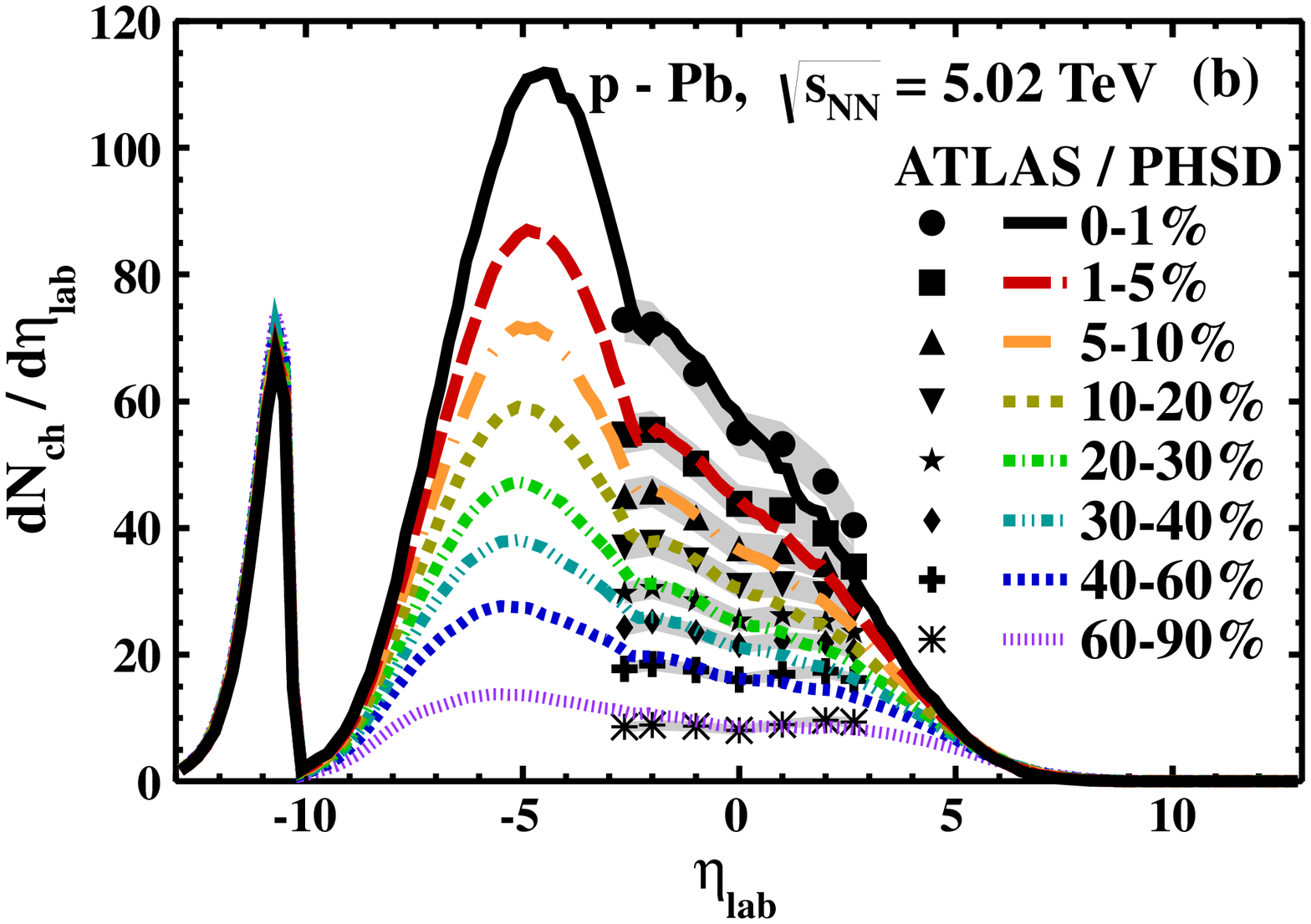}
\caption{(a) Centrality bins for p-Pb collisions at
  $\sqrt{s_{NN}}=$~5.02 TeV selected according to the charged particle
  multiplicity in the rapidity interval $|\eta|<2$. (b) Comparison of
  the PHSD calculated rapidity distributions with ATLAS
  data~\cite{TheATLAScollaboration:2013cja} for charged particles in
  different centrality bins. The shaded bands show the experimental
  uncertainties.}
\label{fig:dndetaS}
\end{figure*}

The CGC predictions, performed earlier for the upcoming p-Pb run at
the LHC, are plotted in the same figure~\cite{Albacete:2012xq} (open
circles). This result is based on the Balitsky-Kovchegov (BK)
equation~\cite{BK96} which is the large-$N_c$ limit of non-linear
renormalization group equations such as the (outlined-above) BK-JIMWLK
hierarchy~\cite{GIJV10} tested with respect to $e+p$ data. The
inclusion of running coupling corrections to the evolution kernel of
the BK equation (rcBK model) made it possible to describe various data
at high energies in terms of solutions of the rcBK
equation~\cite{AK07} and turned out in the best agreement among the
compilation of CGC saturated models in Ref.~\cite{Albacete:2012xq}.
An astonishing result is that the CGC and PHSD results almost coincide
again. Note that this minimum-bias distribution corresponds to the
mean charged particle multiplicity at the given value of
pseudorapidity $\eta$. However, event fluctuations of $dN_{ch}/d\eta$
are very large as demonstrated in Fig.~\ref{fig:dndeta}(b). Thus, the
study of minimum-bias $dN_{ch}/d\eta$ does not allow to disentangle
the initial state concepts described within the PHSD and CGC
approaches.

Let us, furthermore, consider pseudorapidity distributions for fixed
high-multiplicity events. Such distributions for different centrality
bins have been measured recently by the ATLAS
collaboration~\cite{TheATLAScollaboration:2013cja} for p-Pb collisions
at $\sqrt{s_{NN}}=$~5.02 TeV. Experimentally the centrality was
defined according to selected bins in the transverse energy. We have
defined corresponding bins in $N_{ch}$ keeping the same percentage of
the number of selected events as
in~\cite{TheATLAScollaboration:2013cja} (the bin partition is shown in
Fig.~\ref{fig:dndetaS}(a) and the relative contribution of different
centralities is given in the legend in Fig.~\ref{fig:dndetaS}(b)). In
this figure the PHSD results are based on $10^6$ simulated events.

As is seen from Fig.~\ref{fig:dndetaS}(b), the PHSD model quite well
reproduces the shape of the $dN_{ch}/d\eta$ distributions and its
variation with centrality, in particular the increase with centrality
of the forward-backward asymmetry between the directions of the
proton-beam and Pb-target. For the most central events the PHSD
calculations very slightly overshoot this asymmetry, however, are in
line with the data for the higher centralities within the experimental
uncertainties (shaded areas in Fig.~\ref{fig:dndetaS}(b)). We mention
that the centrality sample of 40-60$\%$ with the maximal number
$N_{ch}\sim$~20 roughly corresponds to the minimum-bias
distribution. For events of the highest multiplicity which amount to
(0-1)$\%$ -- corresponding to $\sim 6.10^3$ simulated events -- the
number of charged particles at the maximum of the distribution is
about 75. The agreement between calculations and data is not so bad
taking into account the experimental error bands and the fact that
PHSD has no free parameters once the p-p dynamics is fixed (by the
PYTHIA tune). This holds for p-A as well as A-A reactions in a wide
energy regime and for all centrality classes.

\subsection{Transverse momentum spectra}

\begin{figure}[thb]
\centering
\includegraphics[width=0.49\textwidth]{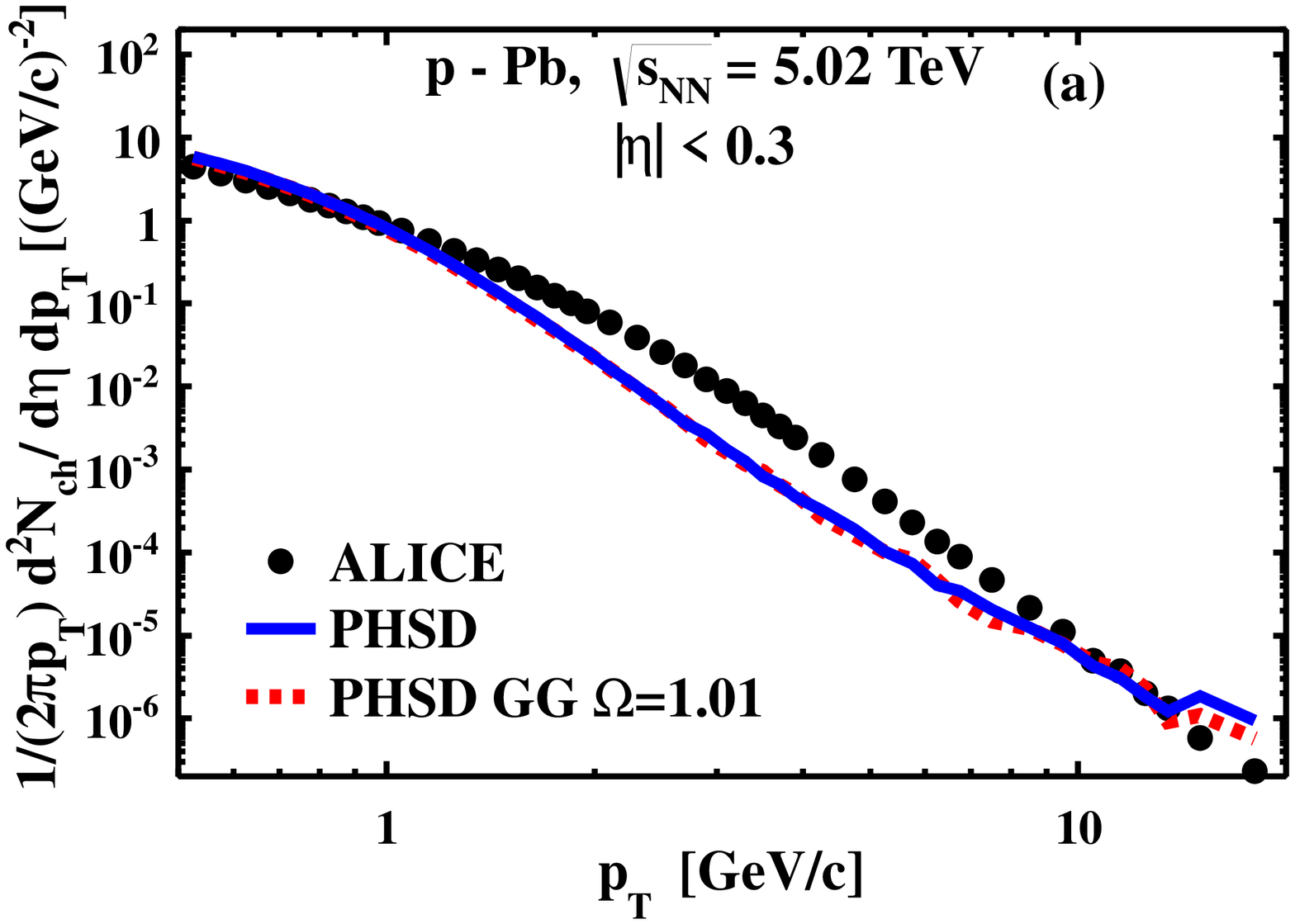}
\includegraphics[width=0.49\textwidth]{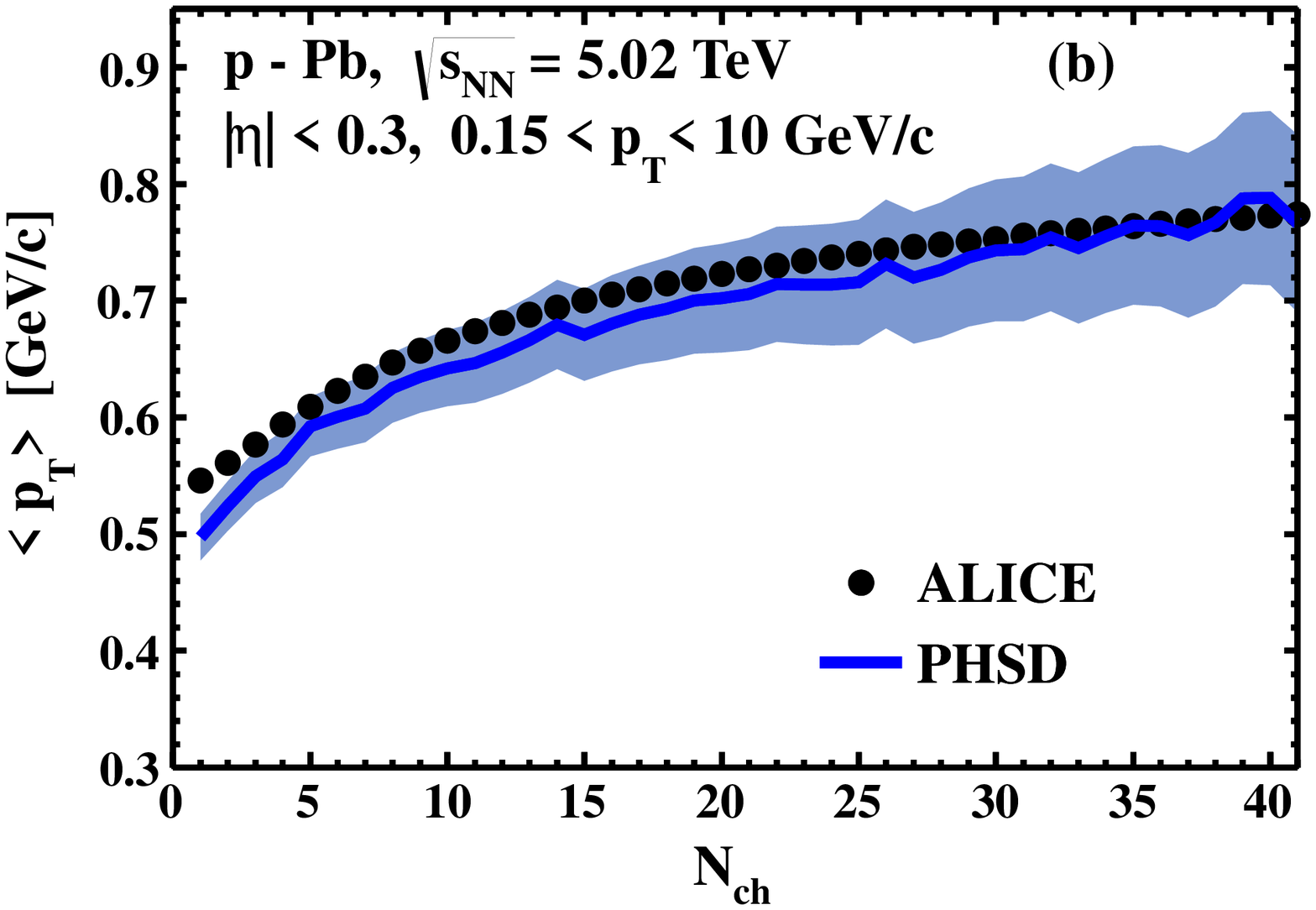}
\caption{(a) Transverse momentum spectrum at midrapidity and (b) the
  mean transverse momentum vs.\ charged particle multiplicity for p-Pb
  collisions at $\sqrt{s_{NN}}=$~5.02 TeV in comparison to the ALICE
  data~\cite{ALICE:2012mj,Abelev:2013bla}. The shaded area in (b)
  shows the statistical uncertainty of the PHSD calculations.}
\label{fig:spectra}
\end{figure}

\begin{figure}[th]
\centering
\includegraphics[width=0.6\textwidth]{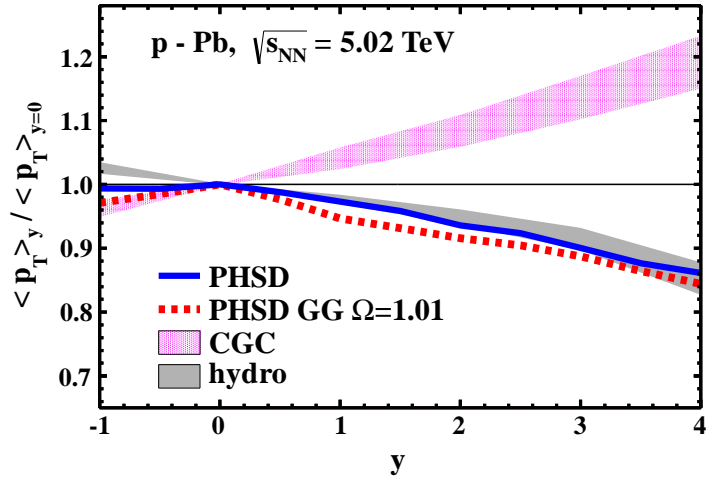}
\caption{The average relative transverse momentum of produced charged
  particles as a function of rapidity for p-Pb collisions at
  $\sqrt{s_{NN}}=$~5.02 TeV. The solid (blue) line is the default PHSD
  prediction whereas the dotted (red) line shows the result including
  large fluctuations in the initial cross section. The CGC and
  hydrodynamic results are taken from Ref.~\cite{BBS13}. The shaded
  areas correspond to the uncertainty in two selected CGC models and
  in the centrality selection in the hydro case, respectively.}
\label{fig:mpt_eta}
\end{figure}

The transverse momentum characteristics for charged particles from the
PHSD for p-Pb collisions at $\sqrt{s_{NN}}=$~5.02 TeV are compared
with the ALICE data in Fig.~\ref{fig:spectra}. In the measured range
$0.5<p_T<10$ GeV/c the yield changes by 7 orders of magnitude in a
rough agreement with experiment~\cite{ALICE:2012mj}. Deviations by up
to a factor of three are observed in the momentum range
$p_T\gtrsim$~1.5 GeV/c (see Fig.~\ref{fig:spectra}(a)) and are
presently not understood. Nevertheless, the dependence of the mean
transverse momentum $\left<p_T\right>$ on the number of charged
particles (Fig.~\ref{fig:spectra}(b)) is rather well described which
implies that the 'soft' physics is sufficiently under control in
PHSD. In view of Fig. 1(d) this result is basically due to the
specific PYTHIA 6.4 tune (390) that rather well reproduces this
correlation for p-p collisions at 7 TeV. We note in passing that
various PYTHIA tunes used before by some experimental collaborations
fail in reproducing this correlation. Additionally, there is
practically no sensitivity to fluctuations in the initial $NN$ cross
sections when using the PHSD-GG version.

A remarkable difference in observables between the predictions of the
saturated CGC and hydro models has been pointed out in
Ref.~\cite{BBS13}. Based on general arguments, it was shown that in
the case of the CGC the mean transverse momentum slightly grows with
increasing rapidity $y$ on the proton side due to the increasing
saturation momentum $Q_s$ of the nucleus (see
Fig.~\ref{fig:mpt_eta}). On the contrary, the
$\left<p_T\right>_y/\left<p_T\right>_{y=0}$ in the hydrodynamical
framework (with Glauber initial conditions) decreases due to the
decreasing number of particles with positive rapidity. This is due to
the fact that the collective expansion scenario (in hydrodynamics)
cannot lead in a simple way to an increase of the average transverse
momentum on the proton side $y >$~0~\cite{BBS13} since there are less
degrees of freedom to generate e.g. a transverse flow. The PHSD model
predicts the $\left<p_T\right>$ distribution (blue solid line) to be
rather close to the hydrodynamic models since also in PHSD the
collectivity is correlated with the density of degrees of freedom
which decreases with forward rapidity (cf. Fig. 5); cross section
fluctuations have no essential influence on this result (dotted red
line). It would be of great interest to check experimentally this
clear difference in the $\left<p_T\right>_y/\left<p_T\right>_{y=0}$
distribution due to different initial state concepts.

\section{Conclusions}

In this study the parton-hadron-string dynamics (PHSD) approach has
been extended to the LHC energy range by implementing additionally the
PYTHIA 6.4 generator (employing the Innsbruck pp tune (390)) to
describe adequately initial hadron interactions in the TeV energy
range (cf. Fig. 1) and to take into account additionally fluctuations
of nucleon-nucleon cross sections (in the sense of the
'Glauber-Gribov' model). The PHSD approach quite reasonably reproduces
observables of p-Pb collisions at $\sqrt{s_{NN}}=$~5.02 TeV, including
those for high multiplicity events, also with respect to the
pseudorapidity dependence. The calculated PHSD results have been
confronted with predictions from saturation CGC models in order to
disentangle the inherent assumptions with respect to the initial state
conditions and dynamics.

We have found that the test for color coherence in the initial state
in ultrarelativistic p-Pb collisions (as proposed in~\cite{BS13})
turned out to be not conclusive. This proposal had been based on
wounded-nucleon model (WNM) estimates of the fraction of high
multiplicity events. However, the WNM does not take into account the
energy-momentum conservation and therefore overestimated this
fraction. Our results within the dynamical PHSD calculations are only
slightly above the CGC predictions and Glauber-Gribov cross section
fluctuations practically do not influence the observables
investigated. Accordingly, the considered quantities, --
multiplicity, average transverse momentum, their distributions and
correlations, -- do not allow for a firm conclusion on the presence
(or absence) of a color glass condensate in a Pb-nucleus at 5.02 TeV.

However, we have found that it is more promising to measure the
$\left<p_T\right>_y/\left<p_T\right>_{y=0}$ distribution (suggested
in~\cite{BBS13}) to obtain a more conclusive result since our PHSD
calculations provide results very close to hydro calculations with a
slope opposite to the CGC models. The physics behind can be expressed
in simple terms: in hydro calculations as well as in PHSD the density
of 'particles' decreases with rapidity on the proton rapidity side
while the saturation momentum $Q_s$ in the CGC increases with $y$.
According experimental studies at the LHC appear feasible and should
allow for a clarification.

\section*{Acknowledgments}
The authors are thankful to Adam Bzdak and Vladimir Skokov for
illuminating discussions. This work in part was supported by the LOEWE
center HIC for FAIR as well as by BMBF. V.~D.\ was also partly
supported by the Heisenberg-Landau grant of JINR.


\end{document}